 \documentclass[reprint,3p,times]{elsarticle}
\usepackage{amssymb}

\usepackage[english,francais]{babel}

\newtheorem{e-proposition}[theorem]{Proposition}

\newtheorem{e-definition}[theorem]{Definition\rm}

\usepackage{amssymb,amsmath}
\usepackage{graphicx}
\usepackage{epstopdf}
\usepackage{color}
\usepackage{url}

\begin{document}
% Select a primary header Physics or Astrophysics
% You can place after the header (classification), if you know it.

\renewcommand{\(}{\left(}
\renewcommand{\)}{\right)}
\newcommand{\D}{\partial}

\newcommand{\w}{\omega}
\newcommand{\FIG}{{\bf\color{red}{FIG: }}}
\providecommand{\un}[1]{\mathrm{#1}}
\newcommand{\REF}{{\bf\color{red}{REF: }}}

\newcommand{\Oi}{\mathcal{O}(\varepsilon)}
\newcommand{\Oii}{\mathcal{O}(\varepsilon ^2)}
\newcommand{\Oiii}{\mathcal{O}(\varepsilon ^3)}

%\centerline{Comtes Rendus Physique: Special issue on phononic crystals}
%\centerline{
%\\Phononic crystals: harnessing the propagation of sound, elastic waves, and phonons}
\begin{frontmatter}


% \author{Name\thanksref{label2}}
\selectlanguage{english}
\title{Nonlinear propagation and control of acoustic waves in phononic superlattices}

% If all authors are at the same address, the [label1] can be suppressed

\selectlanguage{english}
\author[gandia]{No\'e Jim\'enez}
\ead{nojigon@upv.es}
\author[gandia]{Ahmed Mehrem}
\author[gandia]{Rub\'en Pic\'o}
\author[valencia]{Llu\'is M. Garc\'ia-Raffi}
\author[gandia]{V\'ictor J. S\'anchez-Morcillo}
\address[gandia]{Instituto de Investigaci\'on para la Gesti\'on Integrada de Zonas Costeras, Universitat Polit\`ecnica de Val\`encia, Paranimf 1, 46730 Grao de Gandia, Spain}
\address[valencia]{Instituto de Matem\'atica Pura y Aplicada, Universitat Polit\`ecnica de Val\`encia, Cami de Vera s/n, 46022, Valencia, Spain}

% If your know the dates of reception, and acceptation you can put them now;
%    idem the name of the person presenting your article

%\medskip
%\begin{center}
%{\small Received *****; accepted after revision +++++}
%\end{center}

\begin{abstract}
The propagation of intense acoustic waves in a one-dimensional phononic crystal is studied. The medium consists in a structured fluid, formed by a periodic array of fluid layers with alternating linear acoustic properties and quadratic nonlinearity coefficient. The spacing between layers is of the order of the wavelength, therefore Bragg effects such as band-gaps appear. We show that the interplay between strong dispersion and nonlinearity leads to new scenarios of wave propagation. The classical waveform distortion process typical of intense acoustic waves in homogeneous media can be strongly altered when nonlinearly generated harmonics lie inside or close to band gaps. This allows the possibility of engineer a medium in order to get a particular waveform. Examples of this include the design of media with effective (e.g. cubic) nonlinearities, or extremely linear media (where distortion can be cancelled). The presented ideas open a way towards the control of acoustic wave propagation in nonlinear regime.}

\vskip 0.5\baselineskip

%Now keywords/mots-clÈs

\end{abstract}
\end{frontmatter}

% now the Version franÁaise abrÈgÈe, if it exists

\selectlanguage{english}
% main text
\section{Introduction}

One of the most celebrated effects of wave propagation in periodic media is the appearance of forbidden propagation regions in the energy spectrum of electrons, or band-gaps. Most of the physics of semiconductors, and therefore many electronic devices, are somehow based on this concept \cite{Kittel}. In the late 80's, these ideas where extended by Yablonovich and John \cite{Yablonovich87} to light waves (electromagnetic waves in general) propagating in materials where the optical properties like the index of refraction were distributed periodically. These materials were named, by analogy with ordered atoms in crystalline matter, as photonic crystals. The typical scale of the periodicity is given by the wavelength. Actually, not only light but any wave propagating in a periodic medium may experience the same effects, and acoustic waves are not an exception. Sound wave propagation in periodic media has become very popular in the last 20 years in acoustics, after the introduction of the concept of sonic crystals \cite{Sigalas92}. Exploiting the analogies with other type of waves many interesting effects, as the mentioned forbidden propagation bands (band-gaps), but also focalization, self-collimation, negative refraction, and many others have been proposed. We consider in this paper the simplest case plane waves propagating in a 1D structure, formed by a periodic alternation of layers with different properties. Depending on the context, such a structure has been named a multilayer, a superlattice (particularly in the context of semiconductors) or a 1D phononic crystal (this include more exotic structures, as the granular crystal or lattice \cite{Sanchez15}).
  The huge majority of the studies considered so far have assumed a low-amplitude (linear) regime, neglecting the nonlinear response of the medium. Intense wave propagation in nonlinear periodic media, and in particular the case of sound waves, is almost unexplored. In this paper we present different examples of new phenomena related to sound wave propagation in 1D periodic media, where each of the layer has a nonlinear quadratic elastic response. Nonlinear acoustical effects in such structure have been studied only in a few works. In \cite{Yun05} the harmonic generation process is described in a fluid/fluid multilayered structure (water/glycerine), based in a nonlinear wave equation. Also, acoustic solitons in solid layered nonlinear media have been presented in \cite{Leveque03}. More recently, the complementary action of nonlinearity and periodicity has been considered in \cite{Liang09}, where an asymmetric propagation device (acoustic diode) was proposed. There, the nonlinearity and the periodicity act at different locations and its effect is considered separately. 
The effects discussed in this paper are the result of the interplay between nonlinearity and periodicity. Here we describe how the geometrical and acoustic parameters of the structure can be used to control the harmonic distortion processes in a multilayer. The conditions required to selectively act on the nonlinearly generated spectrum, and therefore manipulate the waveform in the desired way, are obtained and discussed.  
	
The theory presented here has been developed for fluid-fluid (scalar) structures, however the main conclusions are extendable to fluid-solid or to solid-solid multilayers, if particular conditions are given. Also, the main conclusions of this paper are independent on the regime of the waves (audible, ultrasound,...), and therefore on the size or scale of the structure. Specially interesting is the domain when ultrasound waves belong to the Terahertz regime, where these ideas may find a great potential. The progress in miniaturization and the technological development allows currently to create phononic multilayers at scales even in the nanometer range (each layer contains then a small number of atoms). This structures are usually made of semiconductors and are often used in particular applications as phononic mirrors to form phonon nanocavities \cite{Huyhn}, or microcavities to obtain a strong optomechanical coupling \cite{Fainstein} (for a revent survey, see \cite{Huyhn2}). In a remarkable recent achievement, acoustic amplification was realized in doped GaAs/AlAs superlattices, where a SASER (Sound Amplification by the Stimulated Acoustic phonon Radiation) was demonstrated, in a device including a superlattice gain medium and GaAs/AlAs SLs acoustic mirrors \cite{Maryam}.

The structure of the paper is as follows: In Sec. 2 we present the model for nonlinear propagation of acoustic waves in periodic media. The next Sec. 3 describes the process of harmonic generation in homogeneous media, and how it is modified by the presence of periodicity. In Sec. 4 the possibility of manipulating the spectrum of a propagating sound wave by tuning the parameters of the layered medium is discussed, showing examples of particular situation, as the case of a cubic-effective medium made out of quadratically nonlinear layers. Finally, Sec. 5 presents the conclusions.

% %%%%%%%%%%%%%%%%%%%%%%%%%%%%%%%%%%%%%%%%%%%%%%%%%%%%%%%%%%%%%%%%%%%%%%%%%%%%%%%%%%%%%%%%%%%%%%%%%%%%%%%%%%%%%

\section{The model}

\subsection{The medium and its dispersion relation}

\begin{figure}[t]
	\centering
	\includegraphics[width=0.6\textwidth]{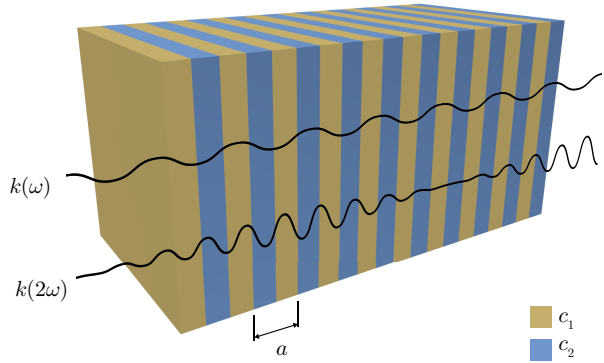}
	\caption{Layered acoustic system with two different layers and second harmonic generation scheme. Here the lattice period is $a=a_1+a_2$.}
	\label{cl:fig:model}
\end{figure}

	We consider a periodic medium made of an arrangement of homogeneous fluid layers of thickness $a_1$ and $a_2$ with different material properties. For the shake of simplicity only longitudinal waves under normal incidence are considered. A scheme of the medium is shown in Fig.~\ref{cl:fig:model}.

\begin{figure}[t]
	\centering
	\includegraphics[width=6cm]{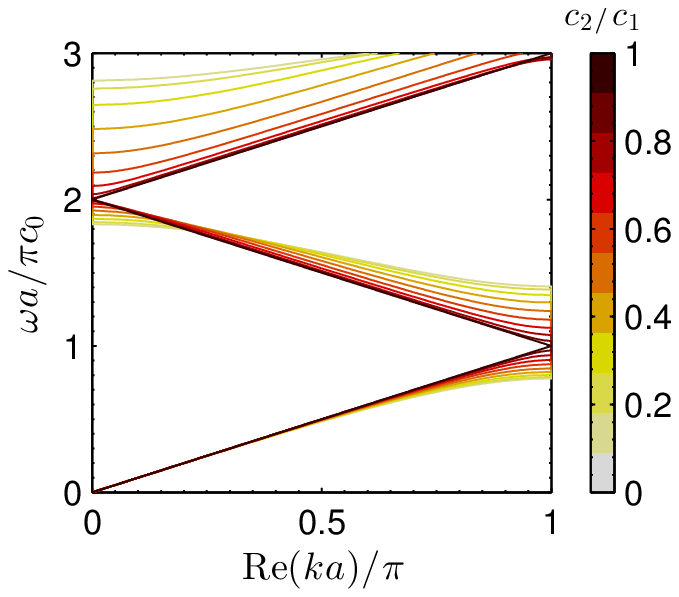}
	\includegraphics[width=6cm]{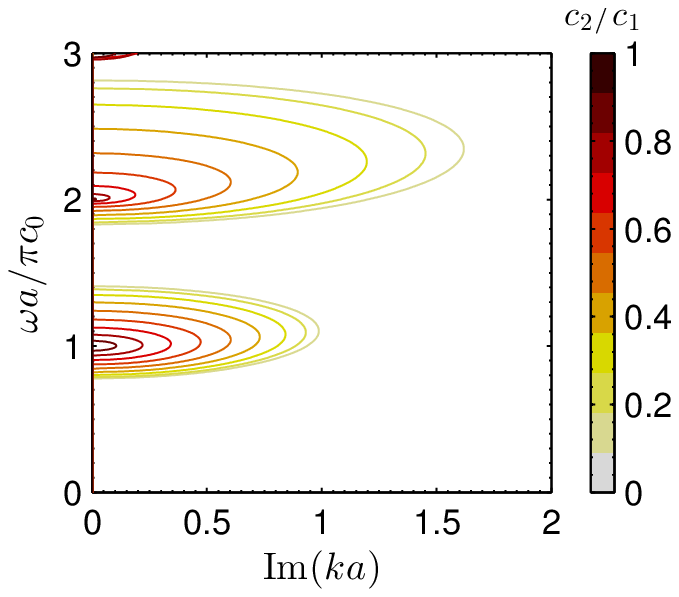}
	\caption{Dispersion relation of the two-layers system for layer proportion $\alpha=0.5$ and for different $c_2/c_1$ ratio. Left: real wavenumbers. Right: imaginary part of the complex wavenumber.}
	\label{cl:fig:dr_sample}
\end{figure}

	The propagation of small amplitude waves in an infinite periodic system is completely described  by its dispersion relation, often known as band structure, that for 1D systems as in Fig.~\ref{cl:fig:model} can be expressed analytically as \cite{Kosevich}

\begin{eqnarray}\label{cl:eq:rytov}
	\cos\left(k a\right)=\cos\left(k_1 a_1\right)\cos\left(k_2 a_2\right)-
	\frac{1}{2} \left(\frac{k_1}{k_2}+\frac{k_2}{k_1}\right)\sin\left(k_1 a_1\right)\sin\left(k_2 a_2\right) \, 
\end{eqnarray}

\noindent also known as the Rytov formula,	where $k$ is the Bloch wave-number, $a=a_1+a_2$ is the lattice period, and $k_i=\omega/c_i$ is the local wavenumber, with $c_i$ the sound speed in the $i$ layer. For a wave of frequency $\omega$ incident in a medium with known acoustical $c_i$ and geometrical $a_i$ parameters, the above equation results in a band structure of propagating and nonpropagating (bandgap) regions, as shown in Fig.~\ref{cl:fig:dr_sample}. Thus, using Eq.~(\ref{cl:eq:rytov}), we can estimate the effect of periodicity on the different harmonics of the incident wave as they propagate through the multilayer, which is the main premise of this work. The ratio between layer thickness can be defined as $\alpha=a_1/a$, leading to $a_2=(1-\alpha)a$.

	An example of dispersion relation plot is shown in Fig.~\ref{cl:fig:dr_sample} for normalized parameters $a=0.5$ and for different sound speed ratios $c_1/c_2$. Increasing the impedance ratio between layers increases the reflected intensity in the trans-layer propagation, while the transmitted energy of the multiple internal reflections diminishes. As can be seen, due to these scattering processes band-gaps are progressively opened around the wavenumber $k=n \pi/a$ with $n=1,2,...$. Thus, the bandwidth of these band-gaps also increases when the impedance ratio grows. 
	
	On the other hand, its imaginary part increases in amplitude with $c_1/c_2$, leading to shorter evanescent propagation inside the band-gap for high sound speed contrast layers, while remains zero (no attenuation) in the propagation band. We recall that the system is conservative: the physical interpretation of the complex wavenumber is not energy absorption, but back-reflection of the incident wave. Thus, at band-gap frequencies the waves penetrate only a short distance into the medium with a forward evanescent mode, and if the medium is perfectly periodic and lossless the energy is back-reflected (it behaves as a mirror).
	
\subsection{Nonlinear constitutive mode}
	The nonlinear propagation of sound in the acoustic inhomogeneous media, and in particular in multi-layered media can be described by several models, with different levels of accuracy. Here, we use the equations of continuum mechanics for ideal fluids with space dependent parameters. These are the continuity equation for mass conservation \cite{Naugolnykh1998}:

\begin{equation}\label{cl:eq:cont}
		\frac {\partial \rho }{\partial t}+\nabla \cdot \(\rho \mathbf{v}\)=0 \,.
\end{equation}

\noindent and the equation of motion that follows from conservation of momentum
\begin{equation}\label{cl:eq:motionD}
		\rho \frac{D \mathbf{v}}{D t} + \nabla p = 0 \,,
\end{equation}

\noindent where $\rho$ is the total density, $\mathbf{v}$ is the particle velocity vector over a Eulerian reference frame, $p$ is the acoustic pressure, $t$ is the time and $D$ is the material derivative operator.
	
	For non homogeneous media, the ambient properties of the fluid in the absence of sound are space dependent, so the total density becomes $\rho(t,{\bf x})=\rho '(t,{\bf x})+\rho _0({\bf x})$, where $\rho _0({\bf x})$ is the spatially dependent ambient density and $\rho'(t,{\bf x})$ is the perturbation of the density or acoustic density, that is space and time dependent. Then, using the material derivative, Eq.~(\ref{cl:eq:motionD}) becomes 

\begin{equation}\label{cl:eq:motionexp}
		\rho _0 \frac{\partial \mathbf{v}}{\partial t} + \nabla p = - \rho' \frac{\partial \mathbf{v}}{\partial t} - \(\rho' + \rho _0\) \( \mathbf{v}\cdot \nabla \) \mathbf{v} \,,
\end{equation}
	
	In this equation, the first two terms in the left-hand-side account for linear acoustic propagation, where the terms in the right-hand-side introduce nonlinearity in the Eulerian reference frame through momentum advection processes.
	
	On the other hand, we can expand Eq.~(\ref{cl:eq:cont}) for nonhomogeneous media as 
	
\begin{equation}\label{cl:eq:contexp}
		\frac {\partial \rho' }{\partial t} + \rho _0 \nabla \cdot \mathbf{v} + \mathbf{v} \cdot \nabla\rho _0 = -\rho' \nabla\cdot \mathbf{v} - \mathbf{v} \cdot \nabla \rho' \,.
\end{equation}	

	Here, the first two terms on the left-hand-side account for linear acoustic propagation, the third, also linear, accounts for the magnitude of the changes in the ambient layer properties. Note this term is space dependent but only changes at the interface between adjacent layers. For density matched layers, $\rho _i=\rho _{i-1}$, this terms vanishes. The terms on the right-hand-side are nonlinear and accounts for mass advection. 
	
	Finally, a fluid thermodynamic state equation $p=p(\rho,s)$ is needed to close the system, with $s$ the entropy. The local nonlinear medium response relating density and pressure variations, retaining up to second order terms, can be written as 
	
\begin{equation}\label{cl:eq:state}
	p=c_0^2\rho ' + \frac{B}{2A}\frac{c_0^2}{\rho _0}{\rho '}^2 \,,
\end{equation}
	
\noindent where $B/A({\bf x})$ is the quadratic nonlinear parameter and $c_0({\bf x})$ is the sound speed, that can be also spatially dependent. 
	
	In this system of equations, quadratic nonlinearity appears in the equation of motion~(\ref{cl:eq:motionexp}) and in the continuity equation~(\ref{cl:eq:contexp}), in the momentum and mass advection terms respectively, and also in the equation of state, Eq.~(\ref{cl:eq:state}), relating pressure and density acoustic perturbations. We note that here we only take into account nonlinear processes through the layer's bulk. The nonlinear effects at the boundary between adjacent sheets are neglected. These nonlinear boundary effects include cavitation processes, that in the case of fluids with very different compressibility can be very important. In the case of solid layers, other local nonlinear effects relative to boundaries, e.g. clapping phenomena between surfaces, can lead to nonlinearities that are orders of magnitude in importance compared to the {\em{bulk}} cumulative nonlinearities.

\subsection{Second-order model}

	For moderate amplitudes, the system of Eqs.~(\ref{cl:eq:motionexp}-\ref{cl:eq:state}) can be simplified. For that aim, we use a perturbative method with same ordering scheme as in \cite{Hamilton2008}, where $\Oi$, $\Oii$ and $\Oiii$ represents the terms of generic smallness parameter $\varepsilon$. The derivation of a second-order nonlinear wave equation requires the substitution of the linearized acoustic approximations (first order) into second order terms of Eq.~(\ref{cl:eq:motionexp}, \ref{cl:eq:contexp}). This substitution procedure will give third order errors, so the final nonlinear wave equation will be a second order approximation of the full constitutive relations.

	These equations can be combined to form a single nonlinear wave equation valid for nonhomogeneous media up to  second order approximation 
	
\begin{equation}\label{cl:eq:wavesecondLag}
		\nabla^2 p - \frac{1}{c_i^2}\frac{\partial^2 p}{\partial t^2} - \frac{1}{\rho _0} \nabla\rho _0 \nabla p = - \frac{\beta}{\rho _0 c_0^4}\frac {\partial^2 p^2 }{\partial t^2} - \( \nabla^2 + \frac{1}{c_0^2}\frac{\partial^2}{\partial t^2}\) \mathcal{L} \,+\,\Oiii\,. 
\end{equation}

\noindent where we introduced the coefficient of nonlinearity $\beta=1+\frac{B}{2A}$ that accounts for material and mass advection quadratic nonlinearities. It is worth noting here  that the second-order Lagrangian density vanish for plane progressive waves due to the first order relation $p=u c_0 \rho _0$ that leads to $\mathcal{L}=0$. In this case, Eq.~(\ref{cl:eq:wavesecondLag}) simplifies to the well-known Westervelt equation for inhomogeneous media

\begin{equation}\label{cl:eq:wavesecondWest}
		\nabla^2 p - \frac{1}{c_0^2}\frac{\partial^2 p}{\partial t^2} - \frac{1}{\rho _0} \nabla\rho _0 \nabla p = - \frac{\beta}{\rho _0 c_0^4}\frac {\partial^2 p^2 }{\partial t^2} \,+\,\Oiii\,. 
\end{equation}

	In general, the Lagrangian density term can be discarded based on the distinction of cumulative and local nonlinear effects. In this way, for progressive quasi-plane wave propagation in homogeneous media the nonlinear local effects become insignificant in comparison to the nonlinear cumulative effects, where in most practicals situations, beyond a distance of only few wavelengths away from the source local nonlinear effects can be neglected. However, local nonlinear effects can become significant in other complex situations including standing-wave fields and finite amplitude acoustic waveguides. Concerning the layered media, in this work we solve numerically the full constitutive relations, and the effect of the Lagrangian term is shown to be negligible under the conditions of our study.

% %%%%%%%%%%%%%%%%%%%%%%%%%%%%%%%%%%%%%%%%%%%%%%%%%%%%%%%%%%%%%%%%%%%%%%%%%%%%%%%%%%%%%%%%%%%%%%%%%%%%%%%%%%%%%
\section{Harmonic generation in layered media}
	
We will study the response of the layered system for plane-harmonic wave excitation. Then, as sketched in Fig.~\ref{cl:fig:model}, the source is placed at one boundary of the layered system, and the acoustic relevant magnitudes are calculated along space and time. As the wave propagates, cumulative nonlinear effects generate harmonics of the fundamental frequency, $\omega _0$, and due to the multiple scattering processes into the layers, local nonlinear effects also distorts the wave. However, the high dispersion of the layered system have a strong impact on the nonlinear harmonic generation. Dispersion modify the resonance conditions between fundamental and second harmonic wave, and also for other nonlinearly generated higher frequencies. In this way, nonlinear energy transfer efficiency from one component to another is modified in a wide variety of configurations, leading to the possibility of engineering and controlling the nonlinear wave processes by tuning the dispersion relation.

Depending of the frequency of the input wave, different scenarios can be observed, as reported in the following subsections.

\subsection{Nondispersive (Fubini) regime}\label{cl:sec:fubini}
	
	\begin{figure}[tb]
	\centering
	\includegraphics[width=6cm]{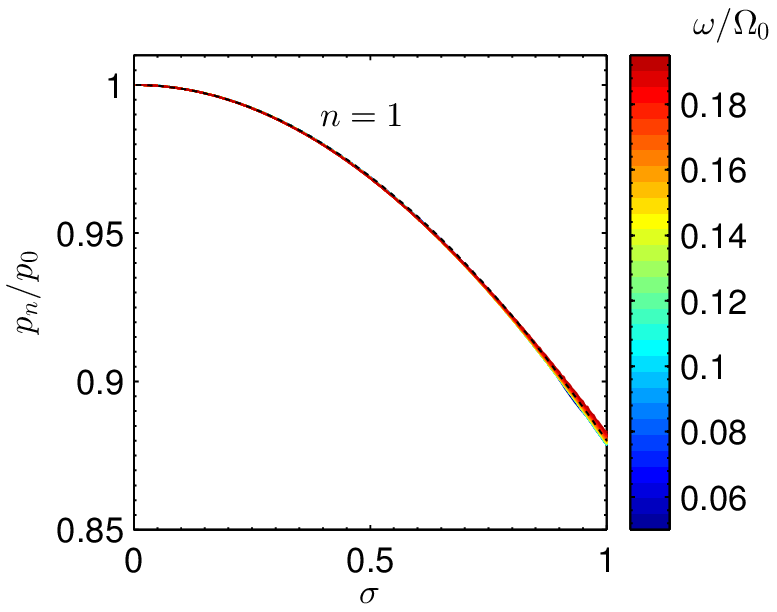}
	\includegraphics[width=6cm]{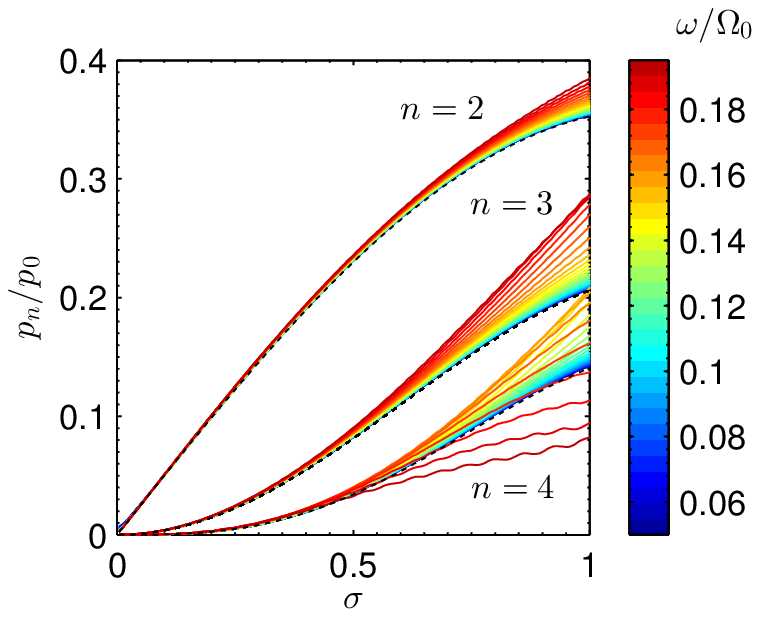}
	\caption{Harmonic generation in the layered medium at low frequencies (numerical results), and its comparison with analytical expressions (Fubini) for an homogeneous medium.}
	\label{cl:fig:distribution_nfc_homogeneous}
\end{figure}

We start studying the propagation in the layered system for harmonic excitation in the very low frequency regime, where we assume that $ka\ll 1 $ holds. As the Rytov's Eq.~(\ref{cl:eq:rytov}) predicts, in the very low frequency regime the slope of the $\omega(k)$ curve is nearly constant. The dispersion of all the spectral components is negligible, and they all propagate at nearly the same velocity and are correspondingly phase-matched. Thus, in the absence of dispersion and attenuation process, the system of Eqs.~(\ref{cl:eq:cont}-\ref{cl:eq:motionD}) and (\ref{cl:eq:state}) can be reduced for a harmonic-plane wave to a Burger's evolution equation expressed in traveling coordinates with effective parameters, namely $\tilde{c_0}$, $\tilde{\rho _0}$ and $\tilde{\beta}$. An analytic solution of this equation in terms of the $n$th-harmonics of the fundamental wave of frequency $\omega$ and initial amplitude $p_0$ is known as the Fubini solution, 
	
\begin{equation}\label{cl:eq:fubinieff}
		p(\sigma,\tau)=p_0\sum _{n=1}^\infty \frac{2}{n\sigma}J_n\(n\sigma\)\sin\(n\omega\tau\), 
\end{equation}
	
\noindent where $J_n$ is the Bessel function of order $n$, and $\sigma=x/x_s$ is the propagation coordinate, normalized to the shock formation distance, $x_s=1/\tilde{\beta} \tilde{\varepsilon} k$, with the effective match number $\tilde{\varepsilon}=u/\tilde{c_0}$ and the effective wavenumber $k=\omega/\tilde{c_0}$, that can be also found from Eq.~(\ref{cl:eq:rytov}). This celebrated solution is valid for $\sigma<1$ (pre-shock region).
	
	Simulations were carried out using a full-wave constitutive relations solver. Thus, we shall define the normalized reference frequency as $\Omega _0=\pi \tilde{c_0} / a$ (located in the first band-gap). The source frequency was set to $\omega=0.1\Omega _0$.
	
	Figure \ref{cl:fig:distribution_nfc_homogeneous} shows the analytical and numerical solutions for the low frequency limit of the layered system, where an excellent agreement is obtained between Fubini and numerical solutions in the pre-shock region, $\sigma<1$ and for low excitation frequencies. As commented above, when the fundamental frequency is increased the higher harmonics fall in dispersive region of the frequency bands, and thus its wave speed is reduced. In this situation, phase matching conditions are no longer fulfilled and therefore the energy transfer from fundamental to higher harmonics is modified. Thus, the Fubini solution can be only applied as an ideal solution for the low-frequency limit or as a good approximation for the first harmonics and for frequencies below $\omega \lesssim 0.1\Omega _0$.

\subsection{Dispersive regime}
	
	For frequencies above the (idealized) homogeneous-Fubini regime, finite (weak and strong) dispersion effects are observed. The dispersive effects of the layered system deeply affects harmonic generation processes. 
	
	As intense waves propagate through a quadratic nonlinear medium, their frequency components interact with each other and new frequencies arise at combination frequencies, including higher harmonics. The cumulative energy transfer from the interacting waves to the harmonics is dependent on the resonance conditions $\omega _1 \pm \omega _2 = \omega _3$, $\mathbf{k}_1 \pm \mathbf{k}_2=\mathbf{k}_3$.
	Note these conditions express the laws of conservation of energy $(\hbar\omega)$ and momentum $(\hbar \mathbf{k})$ in the quantum description for the disintegration and merging of quanta \cite{Naugolnykh1998}. These conditions can be satisfied in a variety of situations. The most simple case is observed in nondispersive media and for collinear waves $k_i=\omega _i/c_0$. In this situation the resonance conditions are fulfilled all over the spectra and a large number of harmonics interacts synchronously: when there exist in the system a \emph{free} wave with velocity $\omega _3 /|\mathbf{k}_3|$ that matches the excited (\emph{forced}) wave $\omega _1 \pm \omega _2 / |\mathbf{k}_1 \pm \mathbf{k}_2|$, the \emph{free} wave is excited in a resonant way. The resonant interaction leads therefore to synchronous (phase matched), cumulative energy transfer from the initial wave to the secondary wave fields.

\begin{figure}[t]
	\centering
	\includegraphics[width=7cm]{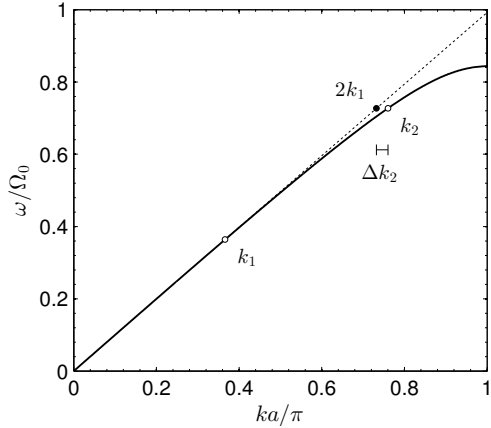}
	\caption{Scheme of the phase miss-matching situation. The fundamental wave vector $k_1$ at frequency $\omega$ generates a \emph{forced} wave $2k_1$ at frequency $2\omega$. The \emph{free} wave that the system allows to propagate is $k_2$, located in the dispersion relation curve. Due to dispersion, $k_2\ne 2k_1$, thus there exist a phase mismatch, $\Delta k_2$ between both waves and the generation is therefore asynchronous.}
	\label{cl:fig:dk_scheme}
\end{figure}	
	
	In the case of an initial monochromatic wave, the main wave generates its second harmonic. The resonant conditions in this situation read $2\omega _1 = \omega _2$, $2\mathbf{k}_1 =\mathbf{k}_2$, that holds true for nondispersive collinear waves, leading to the simple relation $2\mathbf{k}(\omega _1)=\mathbf{k}(2\omega _1)$. However, in the case of dispersive media this condition is, in general, not fulfilled and the \emph{forced} and \emph{free} waves interact asynchronously. Figure~\ref{cl:fig:dk_scheme} shows such situation for a layered media with a fundamental wave in the first dispersion band. 
	
	In order to study asynchronous second harmonic generation processes, we recall here for the lossless second-order wave equation Eq.~(\ref{cl:eq:wavesecondWest}),  for one-dimensional propagation. This equation does not include dispersion by itself, dispersion arises from the solution of the linearized wave equation with the layered media boundary conditions, where the eigenvalue problem leads to the Rytov's dispersion relation Eq.~(\ref{cl:eq:rytov}). 
	
	In the following, we apply a perturbation method to obtain an approximate solution for the second harmonic field. We expand the pressure field as sum of contributions of different orders, i.e. $p= p^{(1)} + \varepsilon p^{(2)} + \cdots$, where $\varepsilon$ is the smallness perturbation parameter, which we identify with the acoustic Match number. Thus, $p^{(1)}$ is the first order (linear) solution of the problem and $p^{(2)}$ its the second order contribution. By substituting the expansion in the second order wave Eq.~(\ref{cl:eq:wavesecondWest}), assuming constant density\footnote{We neglect the ambient density variations for the sake of simplicity. Dispersion arise also for sound speed variations, that are assumed to be implicit in the boundary conditions.} and neglecting $\Oiii$ terms we get a coupled set of equations that can be solved recursively.
	The solution of the first order equation corresponds to a monochromatic plane wave of frequency $\omega$ 
\begin{equation}\label{cl:eq:primarywave}
		p^{(1)}= p_0 \sin\(\omega t - k_1 x\) \,
\end{equation}	
	
\noindent where $k_1=k(\omega)$ is the wave vector associated with the primary frequency $\omega$, and $p_0$ is the excitation pressure amplitude. Substitution of the first order solution into the equation obtained at the next order in the expansion, leads to an inhomogeneous equation for the second harmonic field:
	
\begin{equation}\label{cl:eq:inhomogeneous_second}
	\frac{\partial^2 p^{(2)}}{\partial x^2} - \frac{1}{c_0^2}\frac{\partial^2 p^{(2)}}{\partial t^2} = - \frac{4 \beta \omega^2 p_0^2}{\rho _0 c_0^4}  \sin\(2\omega t - 2 k_1 x\). 
\end{equation}

	The general solution of the this equation is the sum of the solution of the homogeneous equation $(p_0=0)$, and the particular solution of the inhomogeneous equation $(p_0\ne 0)$. Therefore the field for the second harmonic can be expressed as $p^{(2)}=p^{(2)}_h + p^{(2)}_f$, where the corresponding waves for this two solutions are the \emph{free}, and \emph{forced} waves respectively. Such homogeneous and particular solutions are:
	
\begin{eqnarray}
	p^{(2)}_h & = & \left.{p^{(2)}_h}\right\rvert _{x=0} \sin(2\omega _1 t - k_2 x), \label{cl:eq:inhomogeneous_both} \\
	p^{(2)}_f & = & \frac{A}{(k_2 + 2k_1)(k_2 - 2k_1)} \sin\(2\omega _1 t - 2 k_1 x\),\label{cl:eq:inhomogeneous_both2}
\end{eqnarray}

\noindent where $k_2=k(2\omega _1)$ is the wavenumber of the \emph{free} wave at second harmonic frequency, and the constant $A=- {4 \beta \omega _1^2 p_0^2}/{\rho _0 c_0^4}$. It is worth noting here that as long $2k_1\ne k_2$, the \emph{forced} and \emph{free} waves in dispersive media have different phase speed, i. e. the \emph{forced} and \emph{free} waves are phase mismatched as can be seen in the argument of the $\sin$ function in Eq.~(\ref{cl:eq:inhomogeneous_both}-\ref{cl:eq:inhomogeneous_both2}). Imposing the boundary condition, that the second harmonic must be absent at $x=0$, the second harmonic field can be expressed as
	
\begin{equation}\label{cl:eq:evolutionsecond}
	p^{(2)}=\frac{A}{k_2\Delta k} \sin\(\frac{\Delta k}{2} x \) \cos\(2\omega _1 t - k_2'x\),
\end{equation}
	
\noindent where the effective wave number is $k_2'=(k_2+2k_1)/2 \approx k_2$ and the detuning parameter that describes the asynchronous second harmonic generation is defined as
	
\begin{equation}\label{cl:eq:dk}
	\Delta k = k_2 - 2 k_1 = k(2\omega)-2 k(\omega) .
\end{equation}

	Equation~(\ref{cl:eq:evolutionsecond}) describes the well-known effect in second harmonic generation in dispersive media, that is the beatings in space of the second harmonic field when the resonant conditions are not fulfilled. Thus, as $\Delta k$ increases, the beating spatial period and also its maximum amplitude decreases. The position of the maximum of the beating, also called the coherence length, can be related to the second-harmonic phase-mismatching frequency as

\begin{equation}\label{cl:eq:x_c}
	x_c= \frac{\pi}{|\Delta k|} = \frac{\pi}{|k(2\omega)-2 k(\omega)|} .
\end{equation}

	This length corresponds to the half of the spatial period of the beating, where the maximum of the field is located. It can be expressed also for other higher harmonics simply as $x_c(n)= {\pi}/{|\Delta k_n|} = {\pi}/{|k(n\omega)-n k(\omega)|} $.

	In the limiting case of $\Delta k \to 0$, the second harmonic field is generated synchronously and accumulates with distance, so a linear growth is predicted. In this case, phase matching conditions are fulfilled and the \emph{free} wave is excited synchronous to the \emph{forced wave}. Note here that in the derivation of Eq.~(\ref{cl:eq:evolutionsecond}) only second order processes are taken into account and therefore, only second harmonic is predicted. This leads to overestimate the second harmonic field: as long no energy is transferred to third harmonic, second harmonic predicted by Eq.~(\ref{cl:eq:evolutionsecond}) in the absence of dispersion grows indefinitely. The validity of this model can be explored expanding the Bessel functions of Fubini series near the source. A simple comparison between the full Fubini solution and linear second harmonic growth gives a reasonable approximation for distances $\sigma<0.5$ or for second harmonic field values of $p(2\omega)<p_0/4$.
	
\begin{figure}[t]
	\centering
	\includegraphics[width=5cm]{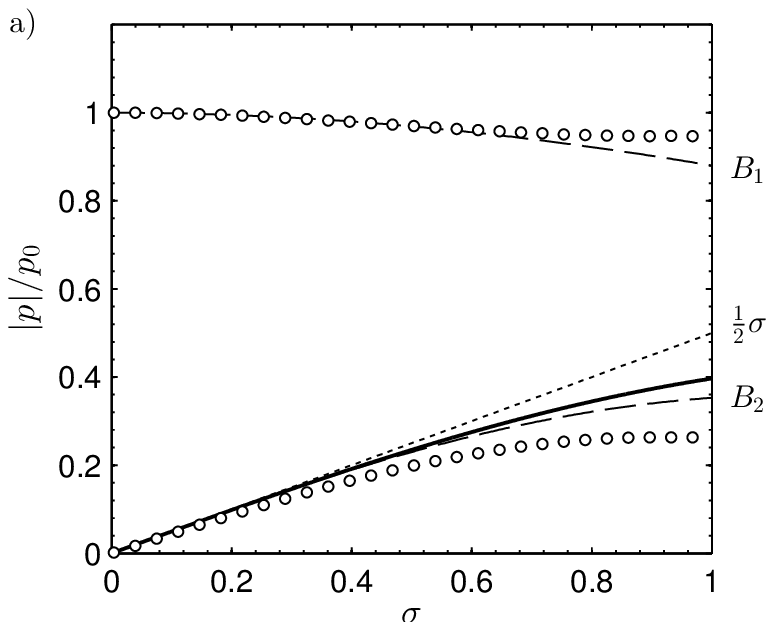}
	\includegraphics[width=5cm]{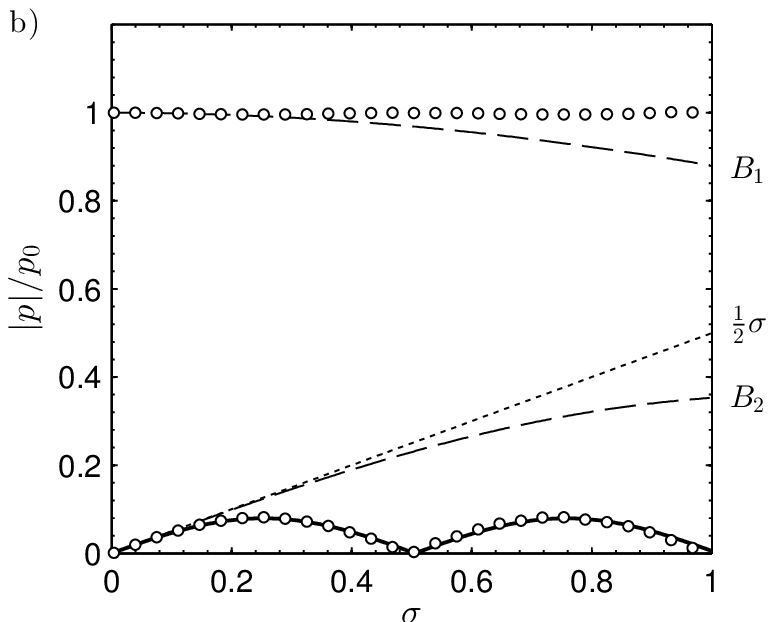}
	\includegraphics[width=5cm]{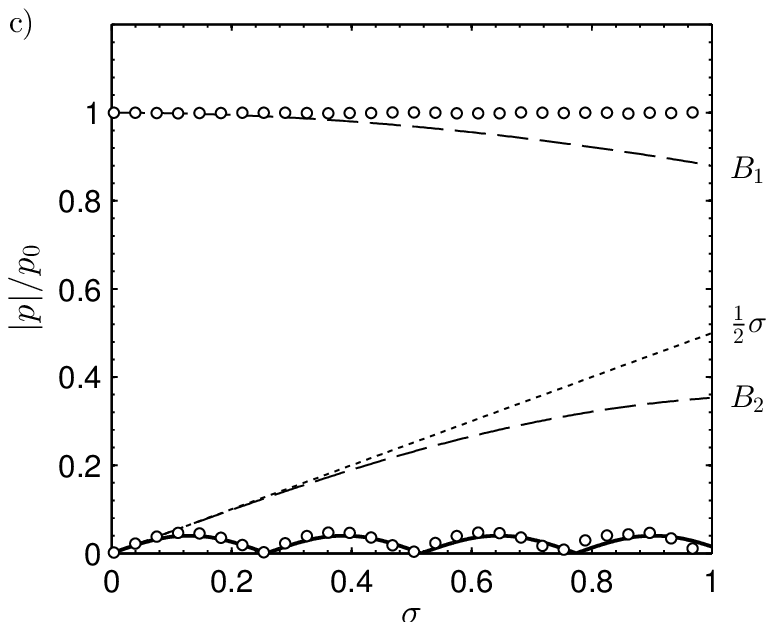}
	\caption{Second harmonic evolution for $x_c/x_s= (1, 1/4$ and $1/8)$ obtained using Eq.~(\ref{cl:eq:evolutionsecond}) (continuous line), numerically (white circles), nondispersive linear law of growth (dotted line) and Bessel-Fubini nondispersive solution (dashed).}
	\label{cl:fig:harmonics_evolution}
\end{figure}	
	
	Figure \ref{cl:fig:harmonics_evolution} shows three different simulations in the dispersive regime of the layered media where the wave amplitude and frequency has been selected to match $x_c/x_s=1, 1/4$ and $1/8$. The higher beating spatial period waves corresponds to lower frequencies. The analytical solution for the second harmonic matches the full-wave numerical solution. However, differences can be observed in the second harmonic amplitude estimation for $x_c/x_s=1$ (Fig.~\ref{cl:fig:harmonics_evolution}~(a). This overestimation by the analytical solution can be related to the absence of energy transfer to higher harmonics, that is not considered by the perturbation solution but is included in the simulation and also in the Bessel-Fubini solution. Therefore, this model is specially suitable in situations where the third harmonic does not grow cumulative with distance. In the lossless layered media, this situations include frequencies that leads to very high-third harmonic detuning and also when the third harmonic falls in band gap.

\subsection{Second harmonic in band gap}\label{cl:sec:SHBG}
\begin{figure}[tb]
	\centering
	\includegraphics[width=5cm]{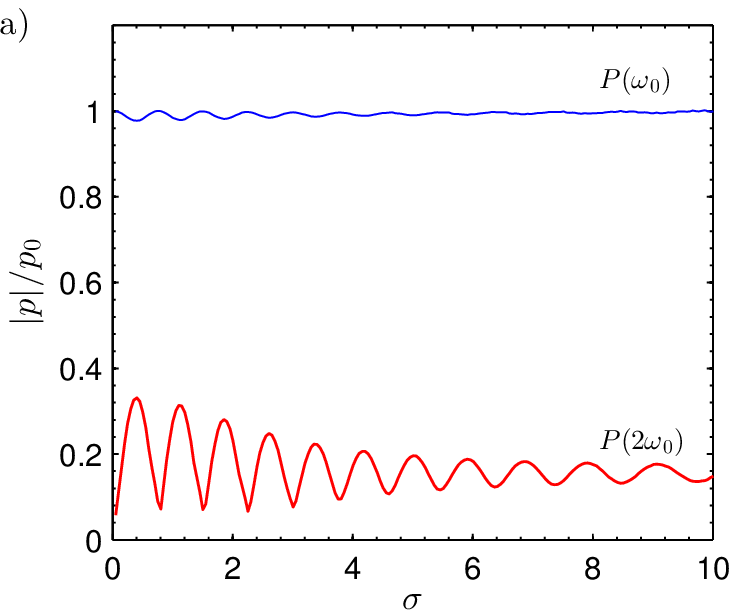}
	\includegraphics[width=5cm]{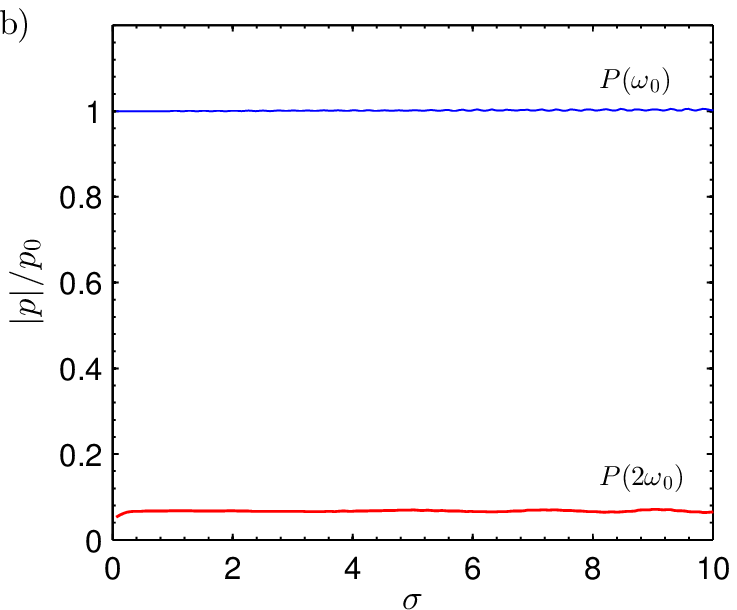}
\includegraphics[width=4.8cm]{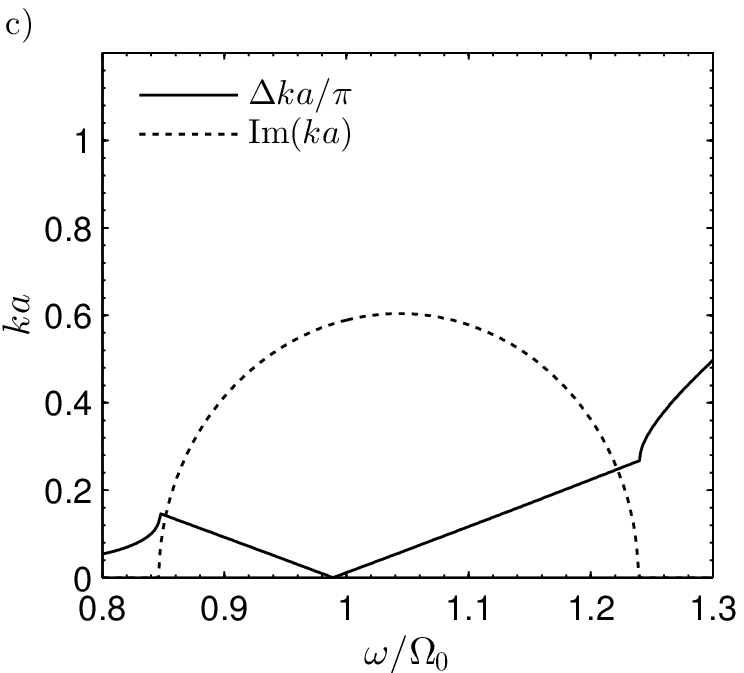}
	\caption{Evolution of the second harmonic field propagating in bang-gap for second harmonic frequencies (a) just above band-gap $2\omega_0=0.84\Omega_0$, (b) in the middle of the bandgap $2\omega_0=\Omega_0$. All results for a layered medium with $\alpha=1/2$ and $c_1/c_2=1/2$. (c) Detuning of the second harmonic (continuous line) and imaginary part (dotted line) in function of the normalized frequency.}
	\label{cl:fig:BG_evolution}
\end{figure}

	Waves with frequencies falling into the band-gap of the dispersion relation are evanescent due the non negligible imaginary part of its complex wave number. Thus, its amplitude decays exponentially with distance. If the nonlinearly generated second harmonic falls into a band-gap, its amplitude does not decay but reaches a constant value \cite{Yun05}. Figure \ref{cl:fig:BG_evolution} shows this case for two different frequencies. The constant amplitude value of the second harmonic wave depends on the imaginary part of the wave vector. 
	
	This effect can be understood in terms of the \emph{free} and \emph{forced} waves. 
	If the second harmonic is evanescent (as follows from the dispersion relation), the wave will not accumulate with distance. The fundamental wave is ``pumping" energy to the second harmonic field at every point in space. Thus, the second harmonic field is generated locally and remains trapped inside the layered media. It reaches a constant level that depends on three main factors. In first place, the "pumping'' rate, characterized by the fundamental wave amplitude and medium nonlinearity, or more strictly the ratio between the layer thickness and the shock distance $a/x_s$. Secondly, it also strongly depends on the magnitude of the imaginary part of the complex wave number, i.e. the ratio between its characteristic exponential decay length and the shock distance in a layer. The characteristic decay length of the evanescent propagation is always shorter when the second harmonic is in the middle of the band-gap, leading to a weaker second harmonic field in this frequency region, as seen in Fig. \ref{cl:fig:BG_evolution}. Finally, it depends also on the detuning of real part of the wave number, where for the first band-gap is minimum at the center. The first factor can be isolated and studied separately. However, the two last factors are linked through the specific dispersion relation of the medium. 
	
	Figure \ref{cl:fig:BG_evolution}(c)~shows the detuning of the second harmonic and the imaginary part as a function of the frequency for a medium with $\alpha=1/2$ and $c_1/c_2=1/2$, showing that at the middle of the band-gap these two factors have opposite effects: detuning is null (phase matching) when evanescent decay is nearly maximized, and viceversa. However, the magnitude of the effects can be very different. As the rate of the second harmonic generation (see the initial slope in Fig.~\ref{cl:fig:harmonics_evolution}) is independent on the detuning, and the evanescence implies that the wave decays after few layers, there not exist a practical compensation of the effects at the center of the band-gap. However, the situation is different for frequencies around the limits of the band-gap, where the coherence length is of the order of the exponential decay characteristic length. Thus, for frequencies just above bad-gap and for amplitudes with shock distance comparable to the evanescent characteristic decay length, the beatings can be also observed, as shown in Fig.~\ref{cl:fig:BG_evolution}~(a). Then, if frequency is increased the characteristic decay length becomes shorter than the shock wave distance and beatings cannot be observed, leading to to the characteristic constant second harmonic field shown in Fig.~\ref{cl:fig:BG_evolution}.

	\subsection{Fundamental harmonic in band gap}
\begin{figure}[t]
	\centering
	\includegraphics[width=6cm]{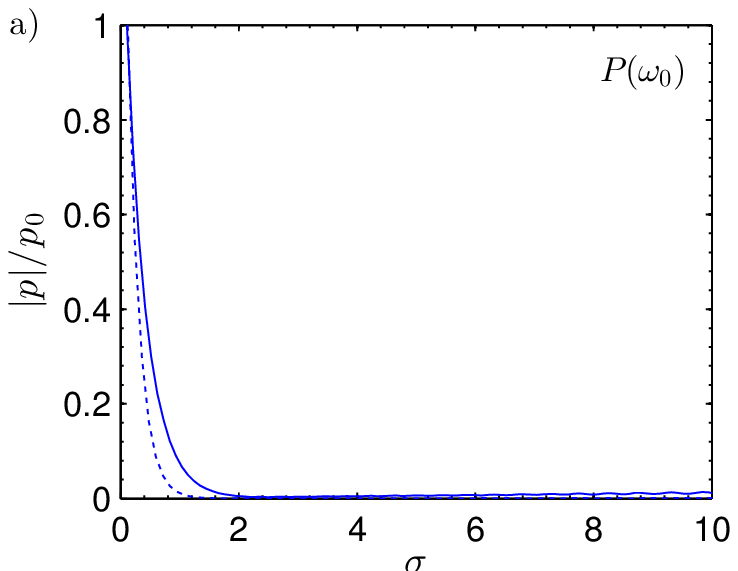}
	\includegraphics[width=6cm]{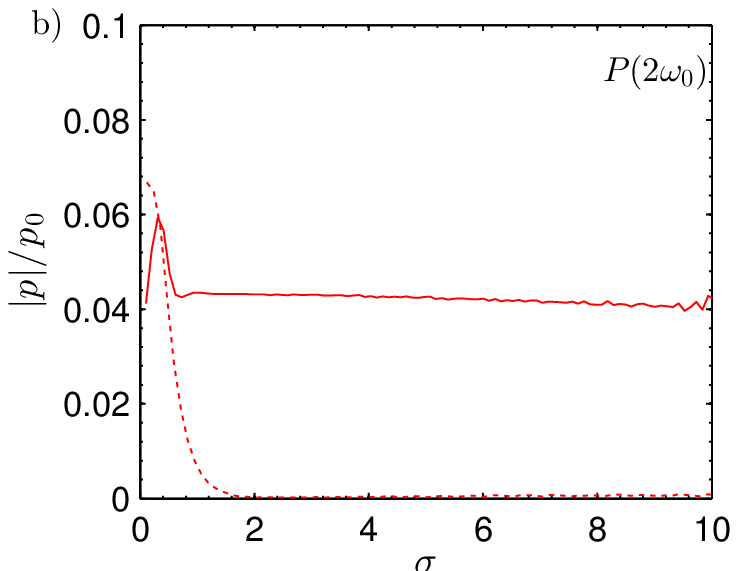}
	\caption{(a) Evolution of the fundamental harmonic wave field with its fundamental frequency falling just above into band-gap, $\omega_0=0.87\Omega_0$, (continuous line), and in the middle of the bad-gap, $2\omega_0=1\Omega_0$ (dotted line). (b) Corresponding second harmonic field, where for $\omega_0=1\Omega_0$ (dotted line) second harmonic frequency falls in the 2nd band band-gap while for $\omega_0=0.87\Omega_0$, (continuous line) lies into a propagating band.}
	\label{cl:fig:BG_first_evolution}
\end{figure}

	When the fundamental frequency of the wave lies within the band-gap, small amplitude waves propagate evanescently. Essentially, the same applies to finite amplitude harmonic waves. In general, if the shock distance is large compared to the characteristic decay length of the evanescent wave, the nonlinear effects have no chance to accumulate and harmonic amplitude is negligible. Since the characteristic exponential decay is about few lattice sites, this means that the initial amplitude necessary to achieve nonlinear effects in this configuration is much higher than those in the preceding sections. Figure~\ref{cl:fig:BG_first_evolution} shows the evolution of the first and second harmonic waves for a fundamental frequency at the Bragg frequency, $2\omega_0=1\Omega_0$, and with a frequency just above but into the band-gap, $\omega_0=0.87\Omega_0$ for a layered media of $\alpha=1/2$ and $c_1/c_2=1/2$. In the first case, the imaginary part of the wave vector is remarkable high and the waves decay fast after few lattice units. Due to this fast decay, the second harmonic interacts only over a short distance with the first, and its amplitude is very limited. After a few lattice units, the fundamental wave can be treated as a small-amplitude evanescent-wave. The second harmonic, that also falls in bandgap (but in the second band gap) also decays exponentially.
	
	On the other hand, if the fundamental frequency is set just above the band-gap, where the imaginary part of the wave-vector is smaller, the amplitude of the fundamental wave decays more slowly, penetrating deeper into the material. The interaction region with the second harmonic is larger, and nonlinear effects result in a more efficient generation of the second harmonic. Furthermore, as long the different (higher order) bandgaps in the layered media can have different bandwidth, in this configuration at $\omega_0=0.87\Omega_0$ second harmonic does not fall inside a bandgap. Therefore, the generated second harmonic wave at the beginning of the lattice propagates through the medium essentially without amplitude change. Due to the evanescence of the fundamental wave, there is only \emph{forced} wave at the beginning of the medium. Therefore, although in this configuration waves are phase mismatched, beatings are not present: only the \emph{free} wave propagates through the medium. 

% %%%%%%%%%%%%%%%%%%%%%%%%%%%%%%%%%%%%%%%%%%%%%%%%%%%%%%%%%%%%%%%%%%%%%%%%%%%%%%%%%%%%%%%%%%%%%%%%%%%%%%%%%%%%%

\section{Nonlinear acoustic field management}

	\subsection{Tuning nonlinearity with dispersion}
	
	In the preceding sections we have explored the fundamental behavior of nonlinear waves generated inside the layered media. But also, medium parameters can be designed to provide specific conditions. The material parameters can be tuned to get coherence at one frequency of interest, e.g. at one of the harmonics of the fundamental wave, or to get detuning or evanescent propagation at other specific harmonics. Using these mechanisms the layered medium can be used to provide a balance of the harmonic amplitudes, or to obtain specific nonlinear waveforms, providing a control of the nonlinear process inside the medium.

\begin{figure}[b]
	\centering
	\includegraphics[width=8cm]{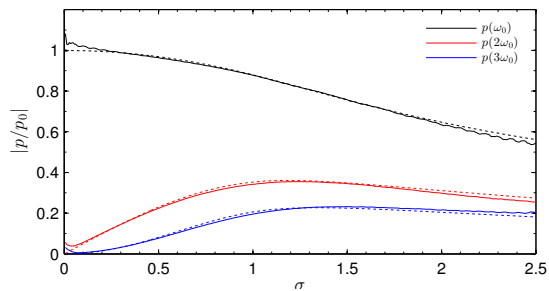}
	\caption{Harmonic distribution for the frequency $\omega_0=1.75/\Omega$. Coherence is recovered for at least the lowest spectral components. Blackstock solution (dotted lines).}
	\label{cl:fig:DR_coherence}
\end{figure}

	In the design of a system for this purpose, the coherence length is a useful control parameter. For this aim, the analytic Eq.~(\ref{cl:eq:rytov}) is used, which is shown to provide an excellent framework to tune the layered parameters to obtain the desired balance between detuning, evanescent propagation, synchronous generation and, at the same time, it allows to find those conditions for a specific phase/group speed. Figure~\ref{cl:fig:increased_second}(a,b) shows an example of a dispersion relation, the coherence length for the second and third harmonic. The resulting harmonic amplitudes when phase matching of all harmonics is achieved is shown in Figure~\ref{cl:fig:DR_coherence}. This happens for a set of frequencies $\omega_0=(0,1.75,2.333,...)/\Omega$. On the other hand, there also exist frequencies at which there exist coherence for the second but a non-negligible detuning is observed for the third. The opposite effect can be also obtained, where coherence is achieved for the third harmonic but second harmonic presents strong dispersion. Finally, other interesting regions are those where second harmonic component is almost phase matched and for the same frequency third harmonic falls into a band-gap.
		
	In the following subsections, we propose and analyze different configurations of the layered medium  with specific balance between  detuning, evanescent propagation and synchronous generation.

\subsection{Enhanced second harmonic generation}
	
	One can expect that second harmonic generation is maximized in homogeneous nondispersive media. However, in nondispersive media coherence is achieved not only at second harmonic frequency, but also in the higher spectral components. As a result, energy is transferred from second harmonic field to higher spectral components and therefore second harmonic field does not grow indefinitely. Moreover, shock waves are formed and nonlinear absorption reduces wave intensity for $\sigma>\pi/2$ even in lossless media \cite{Hamilton2008}. 
	
\begin{figure}[tb]
	\centering
	\includegraphics[width=15cm]{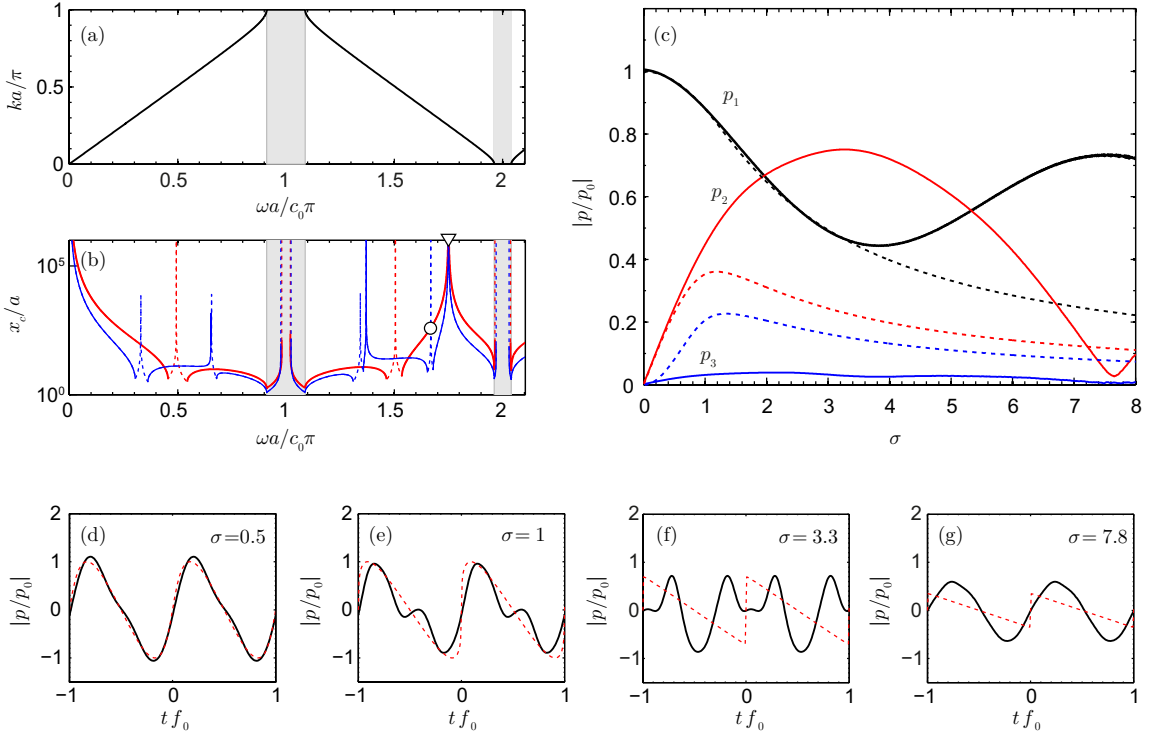}
	\caption{(a) Dispersion relation for a layered media of $c_1/c_2=1.33$ and $\alpha=1/2$. (b) Coherence length for second (red) and third (blue) harmonics as a function of the fundamental frequency. Phase matched frequencies are those with $x_c\to\infty$, while asynchronous generation is predicted for $x_c\to 0$. Frequencies at which the fundamental frequency is in band-gap are are marked in gray regions, while band-gap regions for second and third harmonic are marked in dashed lines.(c) Harmonic distribution for $\omega_0=1.668\Omega_0$ where a coherence is achieved for second harmonic while the frequency of the third harmonic falls into the bad-gap. (d-g) (continuous lines) Waveforms at different distances for $\omega_0=1.668\Omega_0$. At $\sigma=3.3$ second harmonic generation field is maximize and can be seen the period doubling in the waveform. Then, at $\sigma=7.8$ due to second harmonic detuning nearly sinusoidal wave is recovered. 
Analytic Fubini-Blackstock solution for the harmonics (red dotted lines) are plotted for comparison.}
	\label{cl:fig:increased_second}
	
\end{figure}

	The dispersion of the layered system can be used to modify this situation by including phase mismatches that alter the higher harmonic cascade processes, while maintaining coherence for the second harmonic. Figure \ref{cl:fig:DR_coherence} shows an example of a dispersion relation where for $\omega_0=1.668\Omega_0$ it can be observed that there exist a reasonable coherence for the second harmonic ($x_c/a\approx 1000$), while the third harmonic falls in a band-gap. Figure \ref{cl:fig:increased_second} shows the harmonic distribution in this situation. Here, energy is transferred to second harmonic field that grows almost linearly for $\sigma<2$. On the other hand, the energy transferred from second to third harmonic is not cumulative and its amplitude does not grow with distance. Third harmonic experiment evanescent propagation due to the imaginary part of the complex wave-vector at this frequency. A constant field, as studied in Sec.~\ref{cl:sec:SHBG}, is obtained for the third harmonic. 

	The total amount of the second harmonic amplitude in nondispersive media is $p_2|_{\mathrm{max}}\approx 0.36 p_0$, while in the example of Fig.~\ref{cl:fig:increased_second} a maximum second harmonic amplitude of $p_2|_{\mathrm{max}}\approx 0.75 p_0$ is predicted. As can be shown the decreasing of the first harmonic follows the analytic nondispersive Blackstock solution for $\sigma\lessapprox 3$. Thus, in this regime all the energy of the first harmonic is being transferred to the second harmonic field. However, due to finite detuning of the second harmonic a long spatial beating is produced, with normalized period $8\sigma$, and energy is returned back to the first harmonic component. 
	
	It is worth noting here that at distance $\sigma\approx 3$ sawtooth profile is observed in the nondispersive media. In contrast, only second and first harmonic have remarkable amplitude into the layered media. Waveforms are shown in Fig.~\ref{cl:fig:increased_second}(d-g). Near the source, where the amplitude of higher harmonics in not relevant the nondispersive waveform (in red dotted) is well approximated by the fundamental and its second harmonic of the layered medium. However, due to the evanescent propagation of the third harmonic for longer distances the nonlinear solution of the layered medium is mainly composed by the fundamental and its second harmonic. The maximum second harmonic in this configuration is observed at $\sigma=3.3$, as it can be appreciated in the waveforms of Fig.~\ref{cl:fig:increased_second} the period doubling. Moreover, due to finite detuning of the second harmonic the process is not cumulative for all distances and at $\sigma=7.8$ the energy is restored in the first harmonic again and a sinusoidal wave is obtained. Note that not all the energy is restored to the first harmonic in Fig.~\ref{cl:fig:increased_second} at $\sigma=7.8$, leading to a sinusoidal wave of different amplitude as can be observed in Fig.~\ref{cl:fig:increased_second}. The energy loss is mainly due to the artificial (numerical) viscosity necessary to nonlinear convergence\cite{Hamilton2008}. For these simulations the total distance is 1200 lattice sites and therefore the effects of attenuation are not negligible. However, the main nonlinear effects related to strong lattice dispersion still appreciated. An analogous effect has been also studied \cite{Naugolnykh1998} where instead of dispersion, selective absorption at specific frequencies is used to modify and enhance harmonic generation. 

\subsection{Enhanced third harmonic generation}

\begin{figure}[t]
	\centering
	\includegraphics[width=15cm]{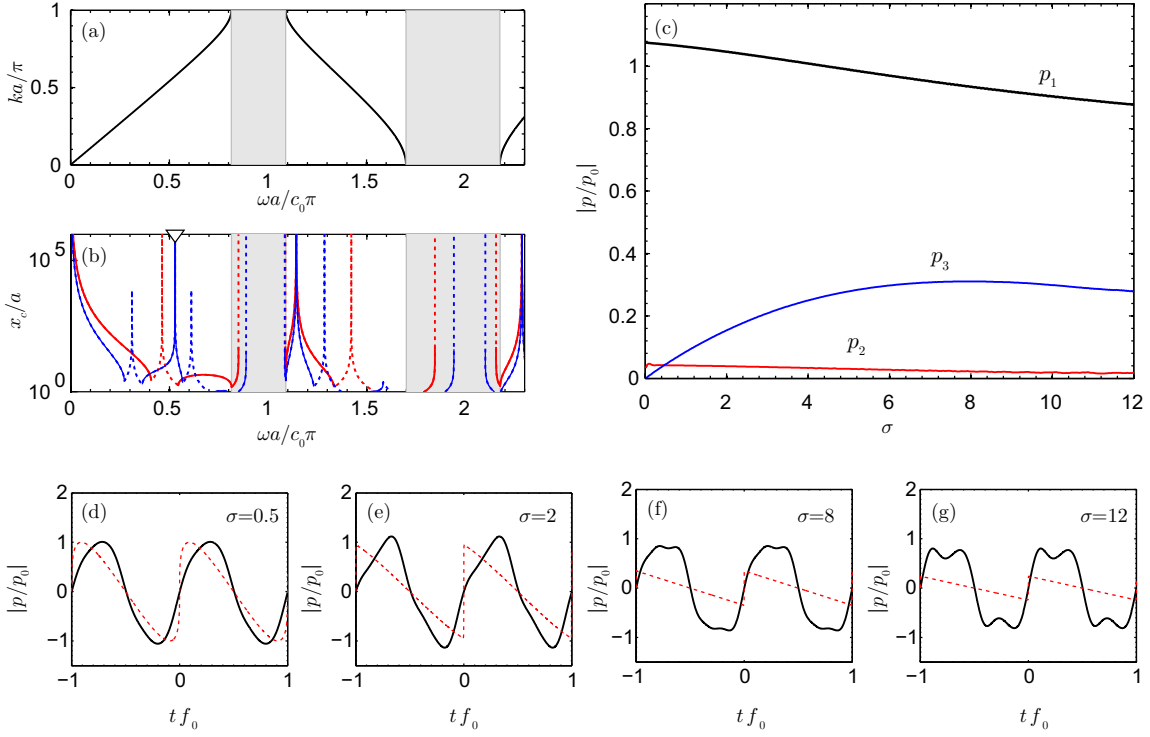} 
	\caption{(a) Dispersion relation for a layered media of $c_1/c_2=1/3$ and $\alpha=0.3$. (b) Coherence lengths for second (red) and third (blue) harmonics, (c) Harmonic distribution for $\omega_0=0.529/\Omega$ where a coherence is achieved for second harmonic while the frequency of the third harmonic falls into the bad-gap. Bottom: (continuous lines) Waveforms at different distances for $\omega_0=0.529/\Omega$. At $\sigma=3.3$ second harmonic generation field is maximize and can be seen the period doubling in the waveform. Then, at $\sigma=7.8$ due to second harmonic detuning nearly sinusoidal wave is recovered. 
Analytic Fubini-Blackstock solution for the harmonics (red dotted lines) are plotted for comparison.}
	\label{cl:fig:increased_third}
\end{figure}

	In the first band ($\omega<\Omega_0$), coherence is always lower for the third harmonic than for the second. However, in the superior bands the layered medium parameters can be tuned to obtain higher coherence for the third than for the second harmonic. Essentially we follow same ideas on the preceding section but for the third harmonic. In this case, the lattice is designed forcing the second harmonic to fall in bandgap. A the same time, perfect coherence can be found for the third harmonic at $\omega=1.4\Omega_0$. This situation is illustrated on Fig.~\ref{cl:fig:increased_third} around $\omega=1.4\Omega_0$. In this case, the dispersion relation was obtained for a layered medium with parameters $\alpha=0.3$ and $c_2/c_1=1/3$.

	In this situation, as Fig.\ref{cl:fig:increased_third} shows, the second harmonic wave attains a constant value of about $0.04 p_0$. As discussed in Sec.\ref{cl:sec:SHBG}, this constant field does not grow with distance and is related to the evanescent solution of the \emph{free} wave and the local nonlinear ``pumping". On the other hand, due to the coherence of the third harmonic, all the energy transferred form second to third is accumulated with distance. Therefore, near the source the rate of energy transfer from second to third harmonic is constant. Thus, third harmonic start to grow almost linearly with distance, opposite to quadratically in homogeneous media. 
	
	Numerical simulations also show fourth and fifth harmonics grow (not shown in Fig.~\ref{cl:fig:increased_third}), but only fifth harmonic harmonic reach a remarkable amplitude, growing near the source almost quadratically with distance. Therefore, the entire system behaves as an artificially cubic-like nonlinear medium formed by quadratic nonlinear layers. 
	
	The corresponding waveforms measured at $\sigma=(0.5, 2, 8, 12)$ are shown in Fig.~\ref{cl:fig:increased_third}(d-g). For $\sigma=0.5$ and 2, it can be observed how the wave steepens with the characteristic shape of cubic nonlinearity. No shock waves are formed as long as strong dispersion is present for high frequency harmonics. It is worth noting here a remarkable fact: it steepens in the positive time axis direction (to the right in the figure), opposite than the quadratic nonlinearity plotted in red dotted as a reference. This effect, i.e. the steepening on the opposite side of the propagation direction, is characteristic of materials with negative parameter of nonlinearity. Therefore, the effective nonlinear behavior observed by the simulations in this conditions can be described as negative-cubic-like nonlinearity.

\section{Conclusions}

 The interplay of dispersion and nonlinearity in multilayered periodic media, such as one-dimensional phononic crystals or superlattices is shown to have a strong impact on the acoustic waves propagating through the structure. Nonlinearly generated harmonics propagating at different velocities are phase-mismatched, modifying the transfer of energy between the different harmonics, and therefore the waveform itself. Shock formation, typical of nonlinear homogeneous media, is in this way avoided. We propose a model and some particular solutions to study this problem, and report examples of configurations that result in an effective control of the spectrum of nonlinear acoustic waves by tuning the dispersion relation of the medium. Selective enhancement of second or third harmonic is demonstrated, leading in some cases to situations where the structure behaves with an effective nonlinearity different from that of its constitutive elements.
 
 The work was supported by Spanish Ministry of Economy and Innovation and European Union FEDER
through project FIS2011-29731-C02-02. A. Mehrem acknowledges Generalitat Valenciana the support from Santiago Grisolia program (grant 2012/029).


\begin{thebibliography}{99}

\bibitem{Kittel} C. Kittel, Introduction to solid state physics. Wiley Eastern Limited (1987).

\bibitem{Yablonovich87} Yablonovitch, E.  Inhibited spontaneous emission in solid-state physics and electronics. Phys. Rev. Lett. 58, 2059 (1987).

\bibitem{Sigalas92} M. M. Sigalas and E. N. Economou, Elastic and acoustic wave band structure, J. Sound Vib. 158, 377-382 (1992)

\bibitem{Sanchez15} Sanchez-Morcillo V.J., Perez-Arjona I., Romero-Garcia V., Tournat V. and Gusev V.E., Second-harmonic generation for dispersive elastic waves in a discrete granular chain. Phys. Rev. E, 88, 043203 (2015)
	

\bibitem{Yun05} Y. Yun, G.Q. Miao, P. Zhang, K. Huang, R.J. Wei, Nonlinear acoustic wave propagating in one-dimensional layered system, Phys. Lett. A. 343, 351-358 (2005)

\bibitem{Leveque03} R.J. Leveque and D.H. Yong, Solitary waves in layered nonlinear media, SIAM J. Appl. Math. 63, 1539-1560 (2003)

\bibitem{Liang09} B. Liang, B. Yuan, and J.-ch. Cheng, Acoustic Diode: Rectification of Acoustic Energy Flux in One-Dimensional Systems, Phys. Rev. Lett. 103, 104301 (2009)


\bibitem{Huyhn2} A. Huynh, N.D. Lanzillotti-Kimura, B. Jusserand, B. Perrin, A. Fainstein, M.F. Pascual-Winter, E. Peronne, A. Lemaître, Phys. Rev. Lett. 97, 115502 (2006).

\bibitem{Huyhn} A. Huynh, B. Perrin, A. Lemaître, Semiconductor superlattices: A tool for terahertz acoustics,  Ultrasonics 56, 66-79 (2015).


\bibitem{Fainstein} A. Fainstein, N.D. Lanzillotti-Kimura, B. Jusserand, B. Perrin, Strong optical? mechanical coupling in a vertical GaAs/AlAs microcavity for subterahertz phonons and near-infrared light, Phys. Rev. Lett. 110, 037403 (2013) .

\bibitem{Maryam} W. Maryam, a.V. Akimov, R.P. Campion, a.J. Kent, Dynamics of a vertical cavity quantum cascade phonon laser structure, Nat. Commun. 4, 2184 (2013) .


\bibitem{Kosevich} Kosevich, A. M. The Crystal Lattice: Phonons, Solitons, Dislocations, Superlattices. John Wiley and Sons (2006).


\bibitem{Naugolnykh1998} K. Naugolnykh and
   L. Ostrovsky, Nonlinear wave processes in acoustics, Cambridge University Press (1998).

\bibitem{Hamilton2008}
M.~F. Hamilton and D.~T. Blackstock, Nonlinear acoustics, Academic Press, San Diego (1998).



\end{thebibliography}
\end{document}